\documentclass[apj, iop]{emulateapj}
\usepackage[usenames]{color}

\usepackage{graphics,graphicx}
\usepackage{natbib}
\usepackage{ulem}
\catcode`\@=11
\def\gsim{\ifmmode{\mathrel{\mathpalette\@versim>}}
    \else{$\mathrel{\mathpalette\@versim>}$}\fi}
\def\lsim{\ifmmode{\mathrel{\mathpalette\@versim<}}
    \else{$\mathrel{\mathpalette\@versim<}$}\fi}
\def\@versim#1#2{\lower 2.9truept \vbox{\baselineskip 0pt \lineskip
    0.5truept \ialign{$\m@th#1\hfil##\hfil$\crcr#2\crcr\sim\crcr}}}
\catcode`\@=12

\definecolor{grey}{rgb}{0.75,0.75,0.75}
\definecolor{Orange}{rgb}{1.0,0.5,0.15}
\definecolor{brown}{rgb}{0.7,0.25,0.0}
\definecolor{pink}{rgb}{1.0,0.5,0.5}
\definecolor{darkerred}{rgb}{0.8,0,0}
\definecolor{darkerblue}{rgb}{0,0,0.8}
\definecolor{Blue}{rgb}{0,0.08,0.65}
\definecolor{Red}{rgb}{0.65,0.08,0.05}
\definecolor{Green}{rgb}{0.15,0.45,0.25}

\newcommand{\kms}{\,{\rm km\,s^{-1}}}
\newcommand{\msun}{\,{\rm M_\odot}}

\newcommand{\bh}{{\rm BH}}

\newcommand{\edd}{{\rm Edd}}
\newcommand{\eps}{{\epsilon}}
\newcommand{\beq}{\begin{equation}}
\newcommand{\eeq}{\end{equation}}
\newcommand{\ba}{\begin{eqnarray}}
\newcommand{\ea}{\end{eqnarray}}
\def\spose#1{\hbox to 0pt{#1\hss}}
\newcommand{\lta}{\mathrel{\spose{\lower 3pt\hbox{$\mathchar"218$}}
      \raise 2.0pt\hbox{$\mathchar"13C$}}}
\newcommand{\gta}{\mathrel{\spose{\lower 3pt\hbox{$\mathchar"218$}}
      \raise 2.0pt\hbox{$\mathchar"13E$}}}

\shorttitle{Supercritical accretion}
\shortauthors{Volonteri, Silk \& Dubus}

\begin{document}

\title{The case for supercritical accretion onto massive black holes at high redshift}

\author{Marta Volonteri\altaffilmark{1,2}, Joseph Silk\altaffilmark{1,3,4} \& Guillaume Dubus\altaffilmark{1,5}}

\altaffiltext{1}{Institut d'Astrophysique de Paris, UMR 7095 CNRS, Universit\'{e} Pierre et Marie Curie, 98bis Blvd Arago, 75014 Paris, France}
\altaffiltext{2}{Department of Astronomy, University of Michigan, Ann Arbor, MI, USA}
\altaffiltext{3}{Department of Physics and Astronomy, The Johns Hopkins University Homewood Campus, Baltimore, MD 21218, USA}
\altaffiltext{4}{Beecroft Institute for Cosmology and Particle Astrophysics, University of Oxford, Keble Road, Oxford OX1 3RH}
\altaffiltext{5}{UJF-Grenoble 1 / CNRS-INSU, Institut de Plan\'etologie et d'Astrophysique de Grenoble (IPAG) UMR 5274, Grenoble, F-38041, France}

\begin{abstract}
Short-lived intermittent phases of  super-critical (super-Eddington) growth, coupled with star formation via positive feedback, may account for early growth of massive black holes (MBH) and coevolution with their host spheroids. We estimate the possible growth rates and duty cycles of these episodes, both assuming slim accretion disk solutions, and adopting the results of recent numerical simulations. The angular momentum of gas joining the accretion disk determines the length of the accretion episodes, and the final mass a MBH can reach. The latter can be related to the gas velocity dispersion, and in galaxies with low-angular momentum gas the MBH can get to a higher mass. When the host galaxy is able to sustain inflow rates at 1--100 $\msun/$yr, replenishing and circulation lead to a sequence of short ($\sim 10^4-10^7$ years), heavily obscured accretion episodes that increase the growth rates, with respect to an Eddington-limited case, by several orders of magnitude. Our model predicts that the ratio of MBH accretion rate to star formation rate is  $10^{-2}$ or higher,  leading, at early epochs, to a ratio of MBH to stellar mass higher than the ``canonical" value of $\sim 10^{-3}$, in agreement with current observations.  Our model makes specific predictions that long-lived super-critical accretion occurs only in galaxies with copious low-angular momentum gas, and in this case the MBH is more massive at fixed velocity dispersion. 

\end{abstract}

\keywords{accretion, accretion disks --- galaxies: high-redshift --- quasars: supermassive black holes}

\section{Introduction} 
The presence of even a few billion solar mass MBHs at high redshift \citep[e.g.,][and references therein]{DeRosa2013} challenges conventional Eddington-limited growth models for MBHs. Assuming Eddington-limited accretion,  a MBH with initial mass $M_0$ grows with time t as $M=M_0\exp\left\{\left[({1-\eta})/{\eps}\right]({ t}/{t_\edd})\right\}$, where $t_{\rm Edd}={\sigma_T \,c}/({4\pi \,G\,m_p})=0.45\,{\rm Gyr}$,  $\eta$ is the fraction of rest mass energy released by accretion, and $\eps \leqslant \eta$ the radiative efficiency, on account of not all of the available energy being necessarily radiated (also jets, winds, see, e.g.,  McKinney et al. 2013 and references therein). In standard radiatively efficient accretion disks $\eta=\epsilon$ and is related to the spin of the black hole with $\epsilon$ ranging from 0.057 to 0.32 for spin parameters ranging from 0 to 0.998.Ê

Given the age of the Universe at $z=6-7$ and the estimated MBH masses, $>10^9\msun$,  {\it constant} Eddington-limited accretion is implied if $10^3<M_0<10^5\msun$ and a duty cycle $\sim$ 50\% for $M_0>10^6\msun$ \citep[e.g.,][and references therein]{Johnson2012}.  From a mathematical point of view, one could imagine resolving the problem by picking the lowest radiative efficiency, $\eps=0.057$.  From the astrophysical point of view, however, the real issue consists in guaranteeing that the host galaxy can  {\it continuously} provide gas, through mergers \citep{2014CQGra..31x4005T} or secular processes, at rates comparable to the Eddington limit for the MBH, despite negative feedback effects \citep{Dubois2013}, and ensure that the MBH is always able to accept this gas. 

\begin{figure}
\includegraphics[width= \columnwidth]{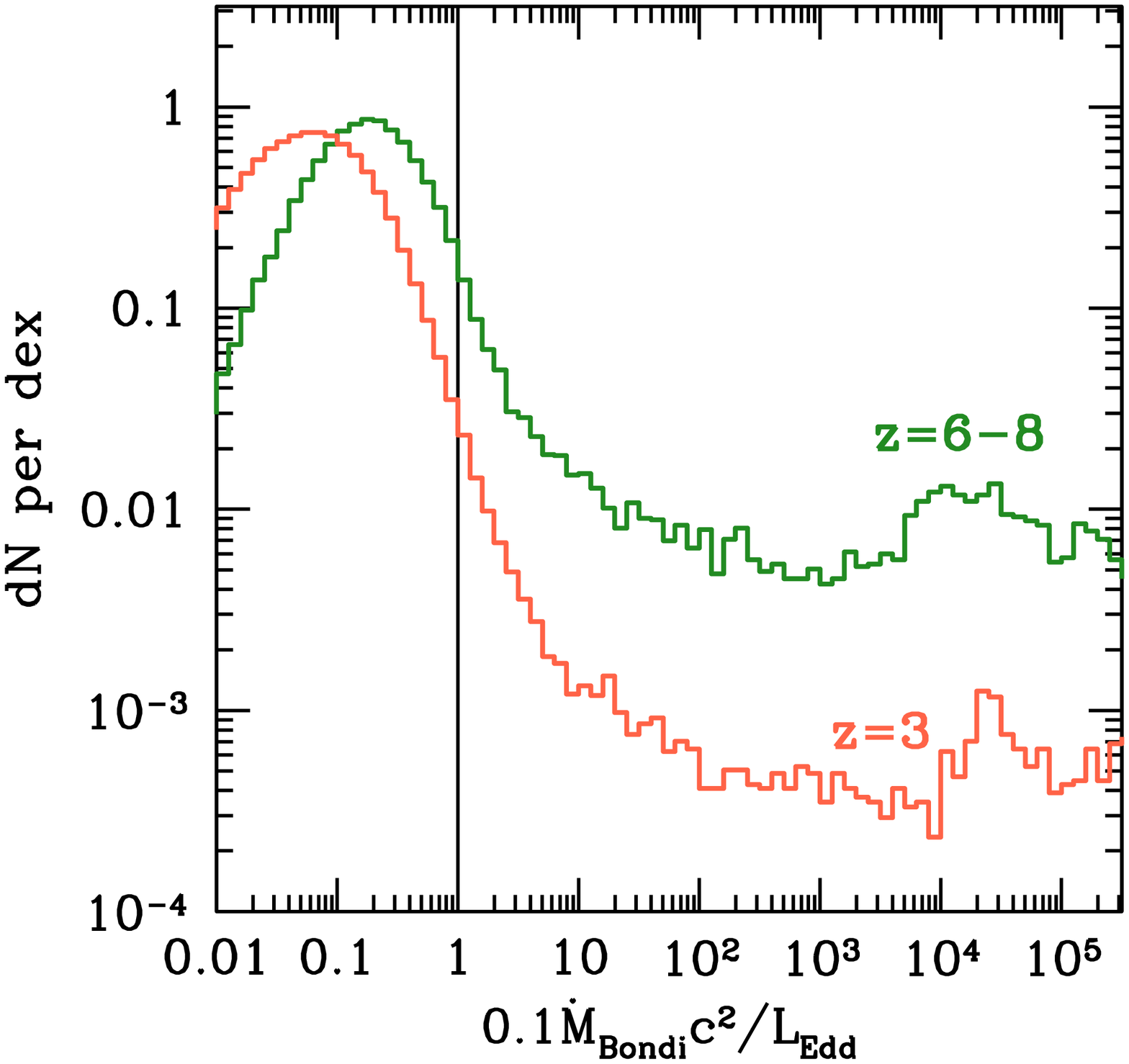}
\caption{Distribution of the ratio between Bondi-Hoyle accretion rate and the accretion rate given by the Eddington luminosity rescaled by the radiative efficiency in the cosmological simulation Horizon-AGN. The fraction of MBHs potentially accreting at super-Eddington rates is $\sim$10\% at $6<z<8$, and it drops to 1\% at $z=3$.} 
\label{fig:horizon}
\end{figure}

In current cosmological simulations of MBH growth, the accretion rate is usually estimated through the Bondi-Hoyle formalism \citep{1944MNRAS.104..273B},  capped at the Eddington luminosity, assuming a given radiative efficiency, normally $\eps=0.1$. However, at the highest redshifts the Bondi-Hoyle accretion rate is often much larger than the limit imposed by the Eddington luminosity.   For illustrative purposes, we show in Fig.~\ref{fig:horizon} the distribution of the ratio between Bondi-Hoyle accretion rate and the accretion rate given by the Eddington luminosity in the Horizon-AGN simulation \citep{2014MNRAS.444.1453D}.  At $z>6$, $\sim$10\% of MBHs in well-resolved halos  experience rates much higher than the Eddington limit;  here only halos resolved with more than a thousand particles, i.e., with $>3\times 10^{10}\,\msun$, are included; MBH masses are $10^5\,\msun$ and above; Dubois et al. private communication. All details about the simulation and the numerical implementation can be found in Dubois et al. 2014).

We stress that the ceiling at the Eddington limit is a general approach used in all cosmological simulations of MBH growth,  both in simulations with a resolution of $\sim1$~kpc (where the Bondi radius is not resolved) and $\sim5$~pc (where the Bondi radius can be resolved), and adopting different codes \citep[smoothed particle hydrodynamics, adaptive mesh refinement, moving mesh, e.g.,][]{2007ApJ...665..187L,2014MNRAS.442.2304H,2014MNRAS.440.1865F,Dubois2013,2014MNRAS.439.2146C}. When the accretion rate surpasses the ceiling, mass growth is  arbitrarily and artificially capped at the rate obtained by imposing the Eddington luminosity with a fixed $\eps$.

 While we should not necessarily take face value the accretion rates from cosmological simulations, the inflow rate of gas may at times be much larger than the MBH Eddington rate, since for a relatively small MBH, e.g., $\sim 10^5\,\msun$ the Eddigton rate corresponds to a small physical accretion rate of $\sim 0.002\,\msun \,{\rm yr}^{-1}$ for $\eps=0.1$, and we explore,  analytically, the consequences for  MBH growth  \citep[see also][]{2012ApJ...749L...3B}, including the consequences of negative or positive feedback originating from the MBH activity  \citep[see also][]{2012ApJ...747....9P}. The former indicates cases where the MBH either suppresses the gas content, thus slowing down or halting the accretion process, or stifles star formation in its surroundings. The latter marks the possibility that a MBH may trigger star formation. We develop arguments in favor of short-lived intermittent phases of  super-critical growth, coupled with star formation via positive feedback, in order to account for early growth of MBHs and coevolution with their host spheroids. 
 
The line of argument of the paper is as follows.  We envisage that when preferentially low angular momentum gas is accreted by MBHs,  the infalling gas will form a very compact accretion disk as rotational support is minimal. Under these conditions initially the disk is under extreme photon trapping conditions, where radiative efficiency and outflows are suppressed. At later times photon trapping becomes less severe, and outflows decrease the net accretion rate, lengthening or stopping the MBH growth timescale.  We show that under such conditions boosts of more than $\sim$ two orders of magnitude to the growth of MBHs with respect to the Eddington-limited case can be achieved. We estimate for how long the Bondi rate can sustain the MBH growth under the envisaged trapping conditions, and we propose that a natural link between the MBH mass and the galaxy gas velocity dispersion is established: in galaxies with much low-angular momentum gas near the center the MBH can get to a higher mass at fixed gas velocity dispersion. Finally, we suggest that jets from rapidly accreting MBHs may account for circulation that replenishes the gas reservoir, and may also explain the high ratio between MBH accretion rate and star formation rate observed in high-redshift quasars.

In this paper we use the following notation:
$L=\eps \dot{M}_{\rm BH} c^2\equiv f_\edd\, L_\edd$;
$t_\edd\equiv{M\, c^2}/{L_\edd}=0.45$ Gyr;
$\dot{M}_\edd\equiv {L_\edd}/{c^2}={M}/{t_\edd}$ (note the absence of $\epsilon$ in this definition);
$\dot{m}={\dot{M}_{\rm BH}}/{\dot{M}_\edd}$, where $\dot{m}$ is a dimensionless ratio of accretion rates, i.e. a normalized accretion rate rather than an accretion rate itself.

\section{Supercritical slim disks}
In general, if material flows through the inner edge of the disk at a rate $\dot{M}_{\rm BH}$, a fraction $\eps$ is radiated away.  It then follows that\footnote{In principle $(1-\eps)/\eps$ should be $(1-\eta)/\eps$. This does not make much difference in the case discussed here since both $1-\eta\sim O(1)$ and $1-\eps \sim O(1)$.}:
\beq
\dot{M}_\bh=\frac{1-\eps}{\eps}f_\edd \frac{M}{t_\edd}=(1-\eps)\dot{m}\frac{M}{t_\edd},
\label{eq:1}
\eeq 
and $f_\edd=\eps \, \dot{m}$. In our terminology, super-critical or super-Eddington accretion rates are defined in terms of the normalized accretion rate, $\dot{m}$, rather than luminosity (in principle, given a sufficiently low efficiency a super-critical MBH may be emitting at sub-Eddington luminosity).


When matter is accreted at intermediate rates ($0.01\lesssim \dot{m} \lesssim 1$) cooling is expected to be efficient and the material forms a geometrically thin and optically thick accretion disk, with the typical solution given by  \cite{shakura&sunyaev73}. At very low accretion rates ($\dot{m} \lesssim 0.01$) cooling becomes inefficient and the forming disk is bloated and radiatively inefficient \citep[e.g.,][]{1995ApJ...452..710N,1995ComAp..18..141A,1999MNRAS.303L...1B}. A MBH accreting super-critically ($\dot{m}\gg1$) is expected to develop a slim accretion disk (Abramowicz et al. 1988, but see Coughlin \& Begelman 2014 for alternatives) with relatively cold temperature and a thick geometric structure.  Slim disks are believed to be radiatively inefficient, being only mildly above the Eddington limit for luminosity, for instance:
\beq
\frac{L}{L_\edd}\sim 2\left[1+ \ln\left(\frac{\dot{m}}{50} \right)\right],
\eeq
for $\dot{m}>50$, and $L/{L_\edd}=\dot{m}/25$ otherwise. We adopted here the expression in \cite{Mineshige2000}, but the logarithmic dependence is a common feature in super-critical disk models, e.g.  \cite{Paczynski1982} and \cite{WangNetzer2003}, and the exact value does not have a strong influence on the arguments presented in this paper.  From $f_\edd=\eps \, \dot{m}$ it follows that the ``effective" radiative efficiency is:
\beq
\eps=\frac{1}{25}  \left(\frac{\dot{m}}{50} \right)^{-1} \left[1+ \ln\left(\frac{\dot{m}}{50} \right) \right],
\label{eq:eps}
\eeq
and $\eps=1/25$ for $\dot{m}<50$. Therefore, while the accretion rate can be highly super-critical, the emergent luminosity is only mildly super-Eddington, because of the logarithmic dependence and low radiative efficiency. Numerical simulations have demonstrated such reduced radiative efficiency at very high inflow rates ($\dot{m}\ga 10^2$, 
\citealt{2005ApJ...628..368O,2007ApJ...659..205O,2011ApJ...736....2O,2013arXiv1312.6127M,2014MNRAS.439..503S}). 

If we assume inflow rates $\dot{M}_{\rm inflow}$ of 1 $\msun/$yr  or 100 $\msun/$yr and an initial MBH mass of $10^5 \msun$, and  self-consistently evolve Eq.~\ref{eq:1} and~\ref{eq:eps}, i.e., we calculate how $\dot{m}$, $M$ and $\eps$ change with time at fixed $\dot{M}_{\rm inflow}$, we find that the growth time would be $\sim 10^7$ yr or $\sim10^{-2}$ of the age of the Universe at $z=6$.

\begin{figure*}
\includegraphics[width= \columnwidth]{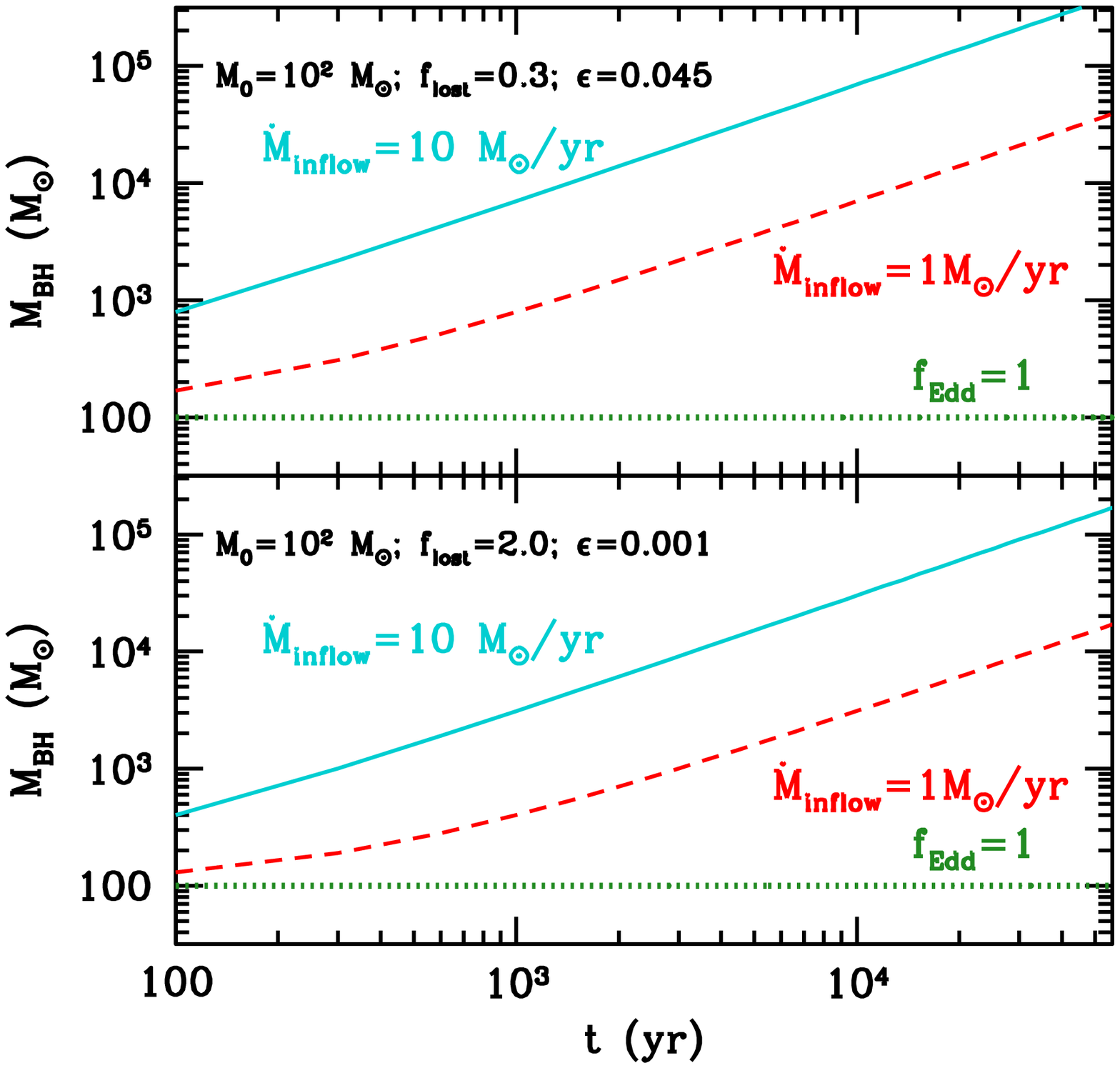}
\includegraphics[width= \columnwidth]{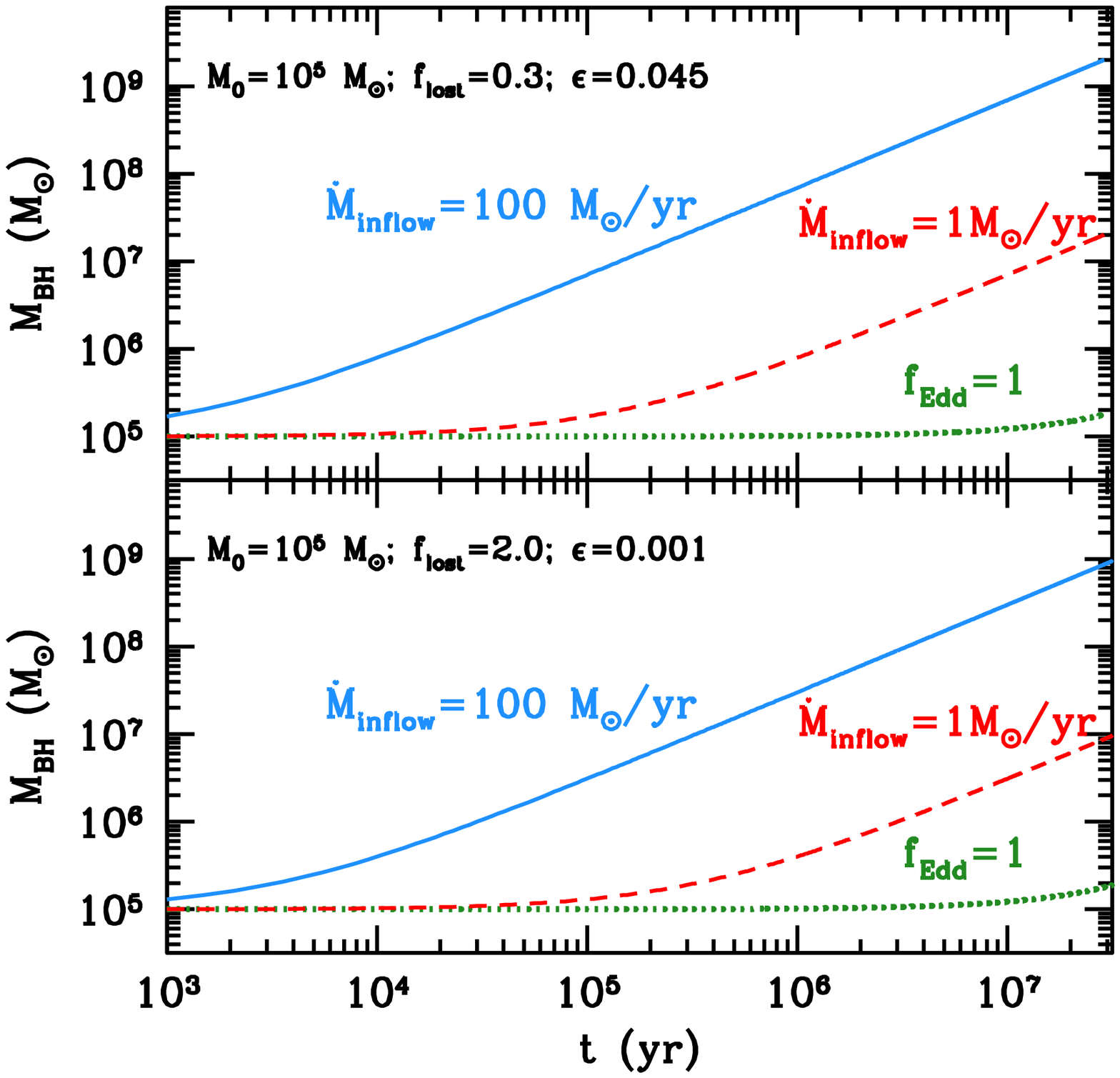}
\caption{MBH mass as a function of time assuming fixed inflow rates from the galaxy $\dot{M}_{\rm inflow}$ of 1, 10  or 100 $\msun/$yr (as marked on the figure) and an initial MBH mass of $10^2$ or $10^5 \msun$ for the radiative efficiencies, $\eps$, and fraction of mass lost, $f_{\rm lost}$, in an outflow found in recent simulations \citep[][see also McKinney et al. 2014]{2014MNRAS.439..503S,2014arXiv1410.0678J}. The MBH growth may be boosted by 2--4 orders of magnitude with respect to the Eddington-limited case (green dotted curve).} 
\label{fig:polish}
\end{figure*}

 The slim disk, however, is a particular solution to the problem of super-critical accretion. Direct simulations may find other solutions, with, for instance, relatively high radiative efficiencies \citep{2014arXiv1410.0678J}, and large outflows that decrease the effective accretion rate on the MBH \citep{2014MNRAS.439..503S}. In the following we assess the consequences of adopting the radiative efficiencies and outflow rates derived in recent simulations.

If there is a significant outflow, and if this outflow carries away most of the inflowing mass then the growth timescale will be longer. We introduce a parameter, $f_{\rm lost}$,  characterizing the mass loss: $\dot{M}=\dot{M}_{\rm inflow}/(1+f_{\rm lost})$.  For instance, \cite{2014MNRAS.439..503S} say in their section 5.4.1 ``as a specific example [for a black hole with spin=0.9], the inflow accretion rate at  30 $R_g$ (where $R_g=2\times R_s$) is $\sim 300\, \dot{m}$. Out of this only $\sim 100\,  \dot{m}$ reaches 10 $R_g$, and the remaining $\sim 200\,  \dot{m}$ goes into an outflow. There is negligible outflow inside 10 $R_g$, so the normalized accretion rate on the MBH is $\sim 100\,  \dot{m}$." In this case the inflow on the MBH would be reduced by a factor of three.   Notably, they find no mass loss for the case of a black hole with spin=0. \cite{2013arXiv1312.6127M} find $f_{\rm lost}\sim1$ for a black hole with spin=0.9375. \cite{2014arXiv1410.0678J} find instead that in their simulation of a non-spinning black hole the mass lost in the outflow is 30\% of the net accretion rate, and $\epsilon=0.045$.  The radiative efficiency given by Eq.~\ref{eq:eps}  is 0.004 at  $\dot{m}=50$ or 0.002 at $\dot{m}=200$, therefore not much different from the 0.0045 derived by Jiang et al. for $\dot{m}=220$.

In Fig.~\ref{fig:polish} we calculate the MBH mass growth at fixed $\dot{M}_{\rm inflow}$ for the $f_{\rm lost}$ and $\epsilon$ found in S{\c a}dowski et al. (2014; $f_{\rm lost}\sim2$ and $\epsilon=0.001$) and Jiang et al. (2014; $f_{\rm lost}\sim0.3$ and $\epsilon=0.045$). \cite{2013arXiv1312.6127M} find $f_{\rm lost}\sim1$ and $\epsilon=0.01$, placing their results in between the two cases shown in Fig.~\ref{fig:polish}.    Even taking into account the mass lost in outflows, and the relatively large radiative efficiency found by \cite{2014arXiv1410.0678J}, this process is much more effective at growing MBHs than Eddington-limited growth.

\section{Radiation Trapping} 
A necessary condition for slim disk accretion is radiation trapping. Trapping of radiation occurs when the photon diffusion time, i.e., the time for photons to escape the disk, exceeds the timescale for accretion. When photons are trapped, they end-up being advected inward with the gas, rather than diffuse out of the disk surface. Plausibly, as long as radiation is trapped in the disk, the emergent luminosity does not exceed much the Eddington limit and the radiative efficiency  stays  low.   The radius at which radiation is trapped can be defined as the locus where the infall speed of the gas equals the diffusion speed of the radiation \citep{1979MNRAS.187..237B}. For $\dot{m}>1$: 
\begin{equation}
R_{\rm tr}=\dot{m}R_s=2\times 10^{-6}\,\dot{m} \,M_7\rm\ pc,
\label{RT}
\end{equation}
where $R_s$ is the Schwarzschild radius of the MBH \citep[this expression may be corrected by a factor $H/R$, of order unity for disks puffed up by the trapped radiation,][]{2002ApJ...574..315O}.  

However, as discussed in \S2, instead of being advected, a significant fraction of the accretion power in supercritical flows may end up driving a disk wind \citep{shakura&sunyaev73,1999MNRAS.303L...1B,2005ApJ...628..368O,2012MNRAS.420.2912B}. Numerical simulations  \citep{2007ApJ...659..205O,2014MNRAS.439..503S} and semi-analytical models \citep{2007MNRAS.377.1187P,2009PASJ...61..783T,2012MNRAS.420.2912B} show that the mass lost to the disk wind is negligible in the inner region and becomes important only as photon trapping becomes less severe (e.g. beyond $\sim 10-100 R_s$ for $\dot{m}\sim 100-1000$ in the references given above).   \cite{2014arXiv1410.0678J} find that an outflow starts in the inner regions in their disk, because of vertical advection driven by magnetic buoyancy, but the outflow appears to be major only in the regions $R \gtrsim 0.1 R_{tr}$. In these regions of the disk  photon trapping is important but not extreme i.e.  the optical depth becomes less than $\sim$10.

Hence, we can reasonably conjecture that a significant disk wind is initiated only after the disk radius has grown to reach some significant fraction of the trapping radius. This has two effects. First, mass lost to the outflow reduces the accretion rate onto the black hole and slows its growth. The simulations indicate the MBH accretion rate never entirely vanishes but can drop to 10-20\% of the inflow rate \citep[e.g.][]{2007ApJ...659..205O}. Second, the importance of the energy and momentum injected in the environment by the disk wind presumably grows with the disk radius, since models and simulations indicate that the wind outflow rate increases as $\dot{M}_{\rm wind}\propto R^s$ ($s\la 1$). The combination of decreasing accretion rate and increased (negative) feedback may eventually quench the black hole growth once the trapping radius is reached \citep[see also][]{VolonteriRees2005}. 

We can take a  general approach by considering the effects of radiation trapping. We can estimate a lower limit to the time-scale for a single accretion episode by requiring that the whole disk's radiation stays trapped i.e. that the outer radius of the accretion disk $R_D\leqslant R_{\rm tr}$.
The accretion disk size is defined in the first place by the centrifugal barrier that determines where infalling material becomes supported by rotation. Let us assume that material is captured near the capture radius of the MBH, $R_g=G\,M/\sigma^2=4.5\,M_7\,\sigma^{-2}_{100}$ pc, where $\sigma_{100}\equiv \sigma/100 \kms$ is the gas velocity dispersion. The material, having angular momentum, does not fall directly into the hole, and it forms an accretion disk. By conserving specific angular momentum one finds that the outer radius of the accretion disk scales with $R_g$, $R_D=\lambda^2 R_g$, where $\lambda \leqslant 1$ indicates the fraction of angular momentum that is retained (it can also signify that only material with angular momentum $1/\lambda$ times smaller than average is captured).

The condition for all radiation to be  trapped, $R_D\leqslant  R_{\rm tr}$, translates into:
\beq
\frac{\lambda^2}{2}\left(\frac{c}{\sigma}\right)^2\frac{M}{t_\edd}\frac{1}{\dot{M}_{\rm BH}}\leqslant 1.
\label{Rd_trap}
\eeq
We can now simplify this expression by approximating $M=\dot{M}_{\rm BH}  t_{\rm acc}$. Rearranging the inequality:
\beq
{t_{\rm acc}}/{t_\edd}\leqslant  {2}{\lambda^{-2}}\left({\sigma}/{c}\right)^2\sim 2\times 10^{-5}\lambda^{-2}_{0.1} \sigma^2_{100},
\label{Dt_trap}
\eeq
For $\lambda=0.1$ and $\sigma=100 \kms$,  $t_{\rm acc} \sim 9000$ yr. With $\dot{M}_{\rm inflow}\sim 100 \msun/$yr, at the end of the episode $M_{BH}\sim 10^{6}\msun$,    and $R_D=R_{\rm tr}\leqslant 0.01$ pc, a value roughly compatible with the half-light radii of accretion disks obtained from microlensing, $\sim 10^{16}$ cm \citep{2011ApJ...729...34B,2015arXiv150105951E}. Hence, with the conservative choice $\lambda=0.1$ (accretion disk sizes favor $\lambda\sim 0.01-0.02$) a MBH would grow $>10^6 \msun$ in $<10^4$ yr. A smaller $\lambda$ would imply a more compact accretion disk, i.e., a smaller $R_D$, and super-critical accretion would proceed for longer, growing the MBH to larger masses.

\begin{figure*}
  \begin{centering}
    \begin{minipage}[t]{0.49\textwidth}
\includegraphics[width= 0.99\columnwidth]{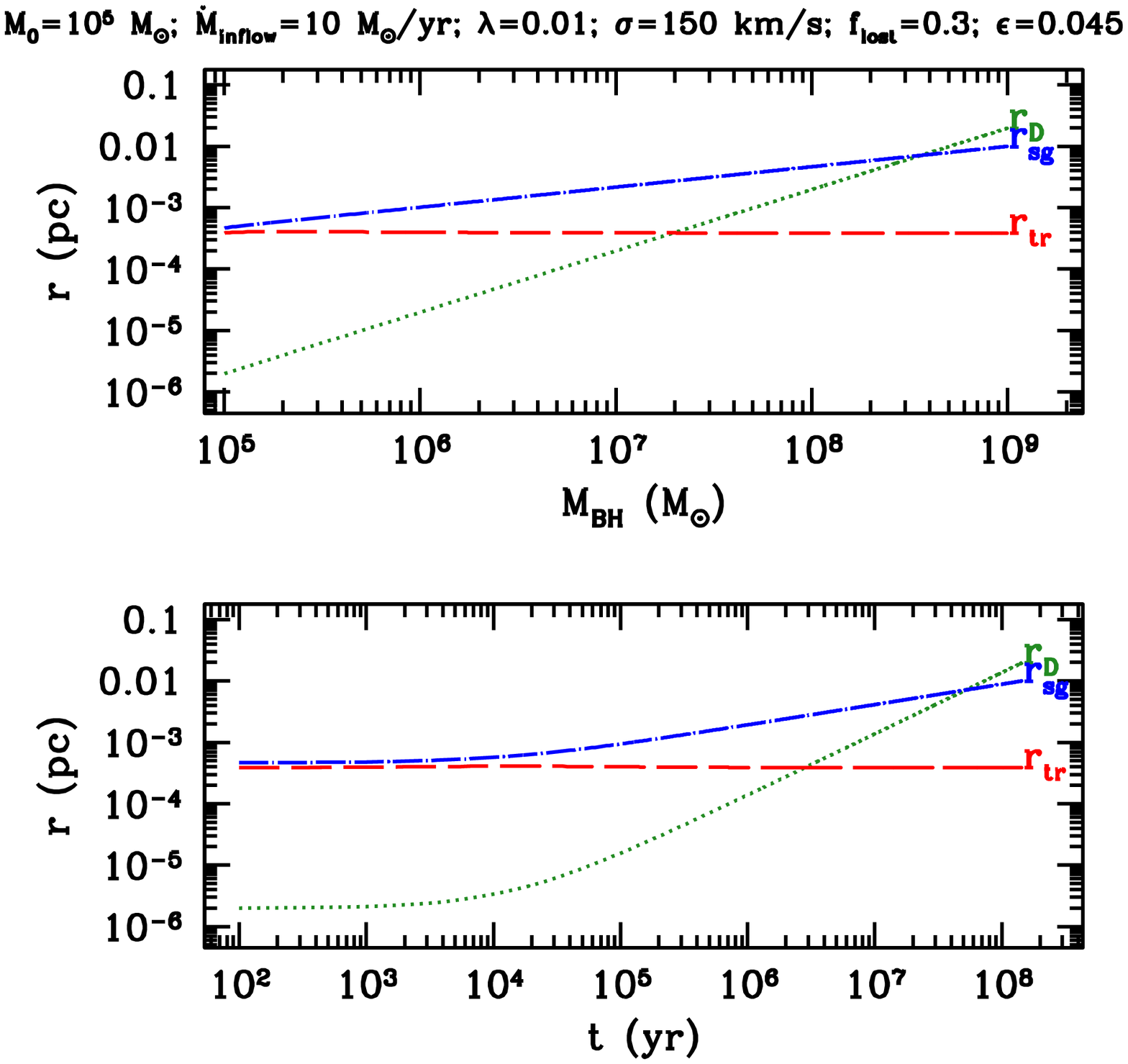}
    \end{minipage}\hfill
    \begin{minipage}[t]{0.49\textwidth}
\includegraphics[width= 0.99\columnwidth]{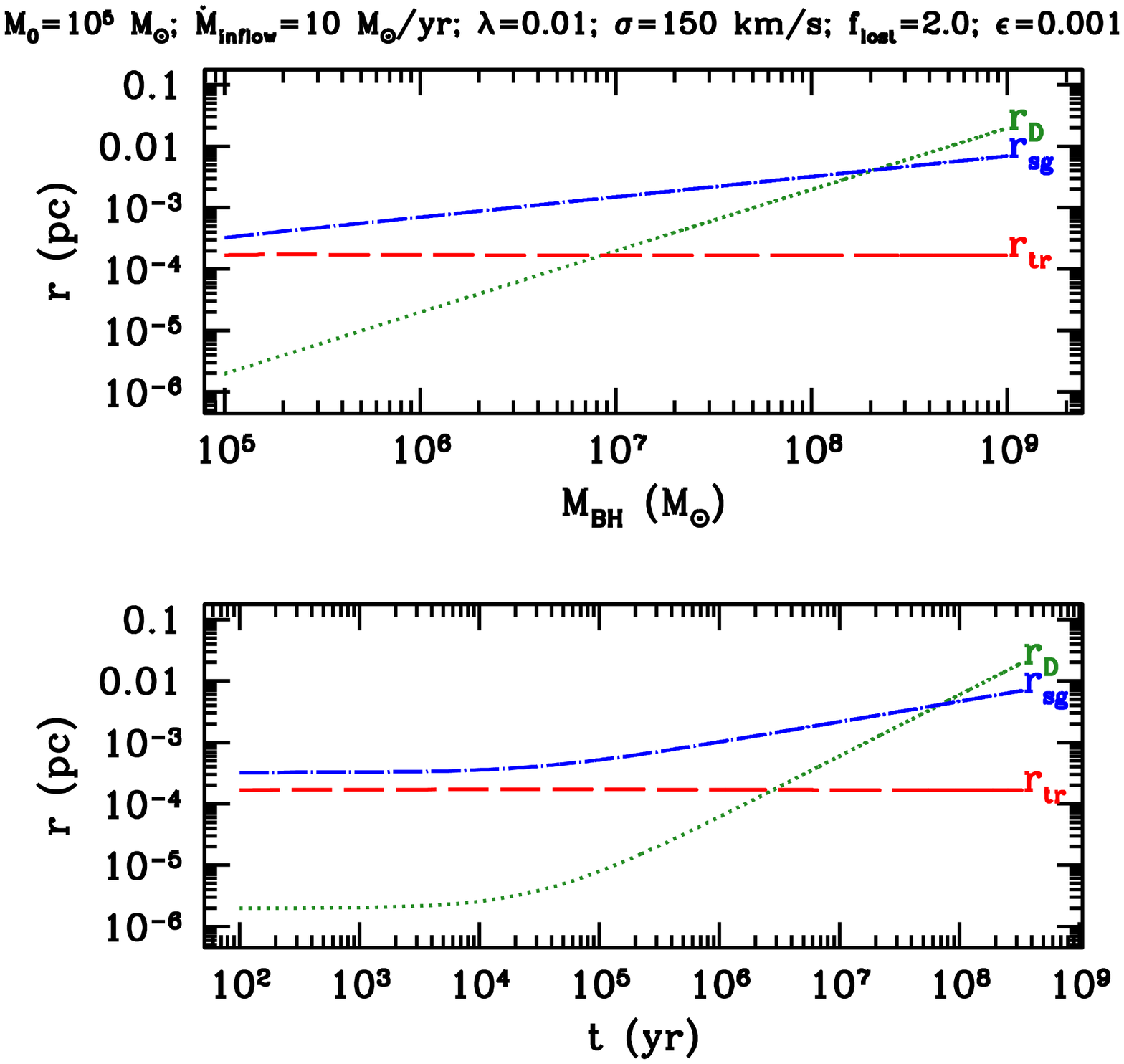}
\end{minipage}
\caption{Disk radius set by centrifugal support ($R_D$, green dotted curve),  self-gravity radius ($R_{sg}$, blue dot-long-dashed curve) and trapping radius ($R_{\rm tr}$, red long-dashed curve) as a function of time (bottom) and MBH mass (top). Growth of a $10^5 \msun $ MBH in a galaxy  with $\sigma=150 \kms$, with a gas inflow rate of 100 $\msun$/yr.  Radiative efficiencies and fraction of mass lost in an outflow as found in recent simulations \citep[][see also McKinney et al. 2014]{2014MNRAS.439..503S,2014arXiv1410.0678J} are marked on the figure. The MBH would grow to $\sim10^7 \msun$ in $\sim 10^6$ years, with a relative increase of a factor $\sim 10^2$ with respect to the Eddington limit. } 
\label{fig:rad}
  \end{centering}
\end{figure*}

 The normalized inflow rates, $\dot{m}$, that we consider are very high: $\dot{m}$ only decreases down to $\sim 10^{4} \lambda_{0.1}^{2}\sigma_{200}^{-2}$ at the end when $R_{\rm D}=R_{\rm trap}$. The initial conditions, with $\dot{m}\gg 1 $  (Eq.~\ref{bondi}) and $R_{\rm D}\ll R_{\rm tr}$, are unlikely to allow for a significant disk wind due to the extreme nature of photon trapping in the flow.   As discussed above, however, even when outflows develop, they may not be able to halt accretion. The duration of a super-critical episode could therefore be longer than we conservatively assumed here.  Under some conditions the accretion disk may be truncated by self-gravity, when the external region of the disk becomes self-gravitating  and subject to fragmentation. For geometrically thin accretion disks \citep{shakura&sunyaev73} the radius beyond which the disk becomes self-gravitating, $R_{sg}$,  can be easily calculated \citep[e.g.,][]{goodman&tan04}, and for slim disks \cite{2004A&A...415...47K} suggest to model them as standard disks outside the trapping radius in order to estimate $R_{sg}$. With this approach, they find that $R_{\rm tr}>R_{sg}$ for $M_7\lesssim 8.8\times 10^6 (\dot{m}/50)^{-2.5} (\alpha/0.01)$, where $\alpha$ is the viscosity parameter. Under the conditions we envisage (i.e., the initial growth of low-mass MBHs with very large $\dot{m}$), the self-gravity radius is typically larger than the radius defined by the angular momentum barrier, as shown in Fig.~\ref{fig:rad}. In this figure we draw the ``nominal" self-gravity radius  for $R_{\rm tr}\sim R_{sg}$ to guide the eye, but we warn the reader that the calculation by  \cite{2004A&A...415...47K}  is valid only when $R_{\rm tr}>R_{sg}$.

 In the scenario we propose, the disk size is initially much smaller than the trapping radius, therefore photon trapping is extreme and we expect most of the inflowing mass being accreted. As the disk grows, photon trapping decreases, and the outflow gains importance until at some point most of the inflowing mass is expelled instead of accreted, stopping the growth process.   In Fig.~\ref{fig:rad}  we model the growth of a $10^5 \msun$ MBH in a galaxy  with $\sigma=150 \kms$, with a gas inflow rate of 100 $\msun$/yr ($\sim$ 12\% of the free-fall rate), and low angular momentum ($\lambda=0.01$). In this case the MBH could grow to $\sim10^7 \msun$, in $\sim 10^6$ years, with a relative boost of a factor $\sim 10^2$ with respect to the Eddington limited growth in the conservative case that accretion is terminated when trapping becomes moderate. The MBH would grow by one additional order of magnitude in the less conservative case that accretion is terminated by self-gravity.

\section{Inflow rate}
One natural question is whether Bondi accretion can provide the huge inflow rate required to sustain the supercritical growth of black holes. The Bondi rate, normalized to $\dot{M}_{\rm Edd}$,  is:
\beq
\dot{m}_{\rm Bondi} \sim \frac{GM c \sigma_T \rho}{m_p \sigma^3}
\eeq
where $\rho$ is the average density of material around the MBH, and $\sigma$ is the gas velocity dispersion. 
To estimate the density, we make here the simplifying assumption of an isothermal spherical distribution.  Since we consider that only low-angular momentum gas ($\lambda \ll 1$) feeds the MBH, the assumption of spherical distribution in the nucleus is plausible. Assuming an isothermal sphere the density at the capture radius becomes $\rho=\frac{\sigma^2 }{2 \pi G r^2} = \frac{\sigma^6}{2 \pi G^3 M^2}$, so

\beq
\dot{m}_{\rm Bondi} = \frac{\sigma^3  c \sigma_T}{2 \pi m_p G^2 M},
\label{bondi}
\eeq
and the trapping condition (cf. Eq.~\ref{Rd_trap}) requires:

\beq
\dot{m}_{\rm Bondi} \geqslant \dot{m}=(1/2) \left({\lambda c}/{ \sigma}\right)^2.
\label{eq:trap}
\eeq

For $\lambda \sim 0.01-0.02$ (based on the size of accretion disks measured through microlensing) and $\sigma\sim 50-200 \kms$ (appropriate for high-z galaxies), the $\dot{m}$ given by Eq.~\ref{eq:trap} is $\sim 100-10^4$, corresponding to $\sim0.01-2 \msun\, {\rm yr}^{-1}$ for a $10^5 \msun$ MBH. The free fall rate ($\sigma^3/G$) in galaxies with $\sigma \sim 50-200 \kms$ would be $\sim 30-2\times 10^3 \msun\, {\rm yr}^{-1}$, therefore the accretion rates necessary for triggering super-critical accretion are a small fraction of the free-fall rate.

Equation~\ref{eq:trap} can be re-written as:
\beq
M\leqslant \frac{\sigma^5  \sigma_T }{\pi \lambda^2 c  m_p G^2} \sim 1.5\times 10^8\, \sigma_{200}^5 \lambda_{0.1}^{-2}\ \msun.
\label{msigma}
\eeq
When the disk radius reaches and exceeds the trapping radius, feedback from the disk in the form of radiation and outflow is likely to become strong enough to blow away the surrounding gas and stop the supercritical accretion phase (see \S~3). The maximum size reached by the disk depends only on $\sigma$ since the disk radius is then equal to the trapping radius, $R_{\rm tr}\sim 0.16\, \sigma_{200}^{3}\,\rm pc$ (combining Eq.~\ref{RT} and Eq.~\ref{bondi}).
At this time a MBH, in a given galaxy with velocity dispersion $\sigma$, has reached a final mass that is modulated by the gas angular momentum, parameterized by $\lambda$: the lower $\lambda$, the longer accretion can continue, the higher the final MBH mass.  Moreover two mechanisms have been suggested that produce considerable angular momentum transfer during the gas-rich collapse phase of the galaxies that we consider, when the MBH underwent much of its growth, the gas was  subject to non-axially symmetric gravitational instabilities. This led to repeated episodes of angular momentum transfer \citep[bar-in-bar,][]{1989Natur.338...45S,2006MNRAS.370..289B}.  Gravitational instabilities also generated massive clump formation, as observed in star-forming galaxies, and dynamical friction  acting on these clumps led to greatly accelerated radial migration of large amounts of gas \citep{2013MNRAS.434..606G}.

This argument allows us to link the MBH mass to the velocity dispersion, and obtain a comparison between the MBH growth and its host properties. Locally,  the relation between MBH mass and velocity dispersion is normalized at $2\times 10^8 \msun$ at 200 $\kms$ \citep[e.g.,][]{2013ApJ...764..184M}, which is close to the relationship derived above (Eq.~\ref{msigma}). 
Galaxies with large amounts of low angular momentum gas ($\lambda\ll1$) will be able to grow MBHs that would appear ``overmassive" at fixed velocity dispersion with respect to the $z=0$ relation. The MBH masses detected in high-z quasars are in fact typically above the $z=0$ relation. We note that the velocity dispersion is estimated for these galaxies through cold gas, rather than stars \citep{Wang2010}, and to derive Eq.~\ref{msigma} we also adopt the gas velocity dispersion. Therefore we conclude that if $\lambda$ is small the MBH can get to a higher mass at fixed gas velocity dispersion with respect to the normalization one would extrapolate from local galaxies (where the velocity dispersion is that of stars. In \S~6 we will discuss  the relative growth of MBH and stellar mass).

In summary, only MBHs hosted in galaxies with copious amounts of low angular momentum gas are able to feed MBHs at supercritical rates for sufficiently long times, by forming small accretion disks where all radiation is trapped for a sufficiently long time.  The importance of low-angular momentum gas is a specific prediction of our model that can be tested with ALMA, at least in principle, via dust continuum observations, on sub-kpc scales.

\section {Circulation and duty cycles}

Jet-driven feedback  or more generally any ultrafast outflow from the Active Galactic Nucleus (AGN) drives a bow shock into the inhomogeneous multiphase accreting gas \citep{2012ApJ...757..136W, wagner13}. The resulting shocks generate  entropy gradients which in turn generate vorticity. A simple derivation of jet interaction with gas clouds that drives circulation of ablated gas is given in
\cite{2010MNRAS.405.1303A}. These crude estimates have been verified in 3-D numerical simulations \citep{2013arXiv1311.5562C}.    Observational evidence for jet-driven back flow is presented in \cite{2012MNRAS.424.1149L, 2014arXiv1402.5109K}. Feedback along the jet direction induces flows that are deflected by the hot spots. The resulting circulation ends up driving accretion onto the disk over the first 1-2 Myr of the cocoon expansion.

Suppose the typical scale, associated with density inhomogeneities in the flow is $L_\omega\sim 0.01L_s,$ where $L_s$ is the size of the system. Turbulent diffusion to larger scales results in circulation that feeds the MBH
along the minor axis of the thick disk. The diffusion coefficient is $\sim L_\omega v/3.$ The time to regenerate the feeding of the MBH is 
$
\sim 3 \left( {L_s/ L_\omega} \right)
 \left({L_s/ v}\right). 
$ 
Plausible numbers suggest a crossing time-scale $\sim L_s/v\sim 10^{2-3} \rm yr,$ and a regeneration time  $t_{\rm reg}\sim 10^{4-5}\rm yr.$

We have argued that each accretion episode lasts $\sim 10^{3-5}$ yr, with the shortest accretion episodes associated to $\lambda=1$ (Eq.~\ref{Dt_trap}). Per accretion episode, the MBH grows in mass by:
\beq
\Delta M=\dot{M}_{\rm BH} t_{\rm acc}=\dot{m} \dot{M}_{\rm Edd}t_{\rm acc}\sim 10^5 \msun \dot{M}_{100} t_{\rm acc,3},
\eeq
where the MBH accretion rate is in units of 100$\msun/$yr  and time in units of $10^3$ yr.
Each accretion episode is followed by a flow regeneration period 
of order $t_{\rm reg}\sim 10^{4-5}\rm yr$ based on circulation and replenishment arguments.

Since $t_{\rm reg}>t_{\rm acc}$,  the maximum number of super-critical accretion episodes will be
$
N={t_{\rm Edd}}/{t_{\rm reg}}\sim  10^4 t_{\rm reg,5}^{-1} 
$
for $t_{\rm reg}$ in units of $10^5$ yr.  If at each episode the MBH grows in mass by $\Delta M\sim 10^5\msun$, then the total growth is $N\times \Delta M\sim 10^9 \msun$, and the duty cycle is $\delta=N{t_{\rm acc}}/{t_{\rm Edd}}\sim 0.01\, t_{\rm acc,3}/t_{\rm reg,5}$. However, this assumes each episode lasts $t_{\rm acc}$ whereas subsequent episodes may actually be much shorter. According to the assumptions in the previous sections, the outer radius is necessarily  close to the trapping radius if accretion has already been quenched once. Any resuming accretion may be rapidly re-quenched once the disk reforms on a dynamical timescale. As the MBH grows in mass, the importance of multiple supercritical accretion episodes therefore decreases.

%

\section{MBH accretion and star formation}
 MBHs accreting at  super-critical rates are most likely to be characterized by strongly collimated outflows or jets, especially if the MBH is rotating \citep{2014MNRAS.439..503S}. Such collimated outflows would not cause negative feedback directly on the surrounding gas, which is pierced through \citep{2006ApJ...645...83V}, with only a very small fraction of the gas being directly affected \citep[$\sim$ 1\% according to the recent simulations by][]{2013arXiv1311.5562C}. 

For gas-rich hosts, relevant at early epochs, the AGN  jet (or wind) driven by the MBH can instead trigger positive feedback, and be strongly coupled to the star formation rate, presumably because of bow shock pressure-enhanced star formation \citep{2012MNRAS.425..438G, wagner13}. Evidence for this comes directly from radio-selected samples \citep{2013ApJ...774...66Z} and more indirectly from the presence of massive, remarkably young stellar populations in high redshift radio galaxies \citep {2013MNRAS.429.2780R}.

There are numerous examples of jet-induced positive feedback at low-$z$, where high radio resolution has been useful, two  recent examples being given by 
\cite{2013A&A...552L...4M}  and \cite{2013A&A...558A...5R}. Moreover  recent data \citep{2013arXiv1312.2417M} suggests that the most luminous phase
of the AGN correlates with star formation and most likely precedes  any phase of negative feedback, as motivated observationally by massive 
molecular outflows \citep{2011ApJ...733L..16S,2013arXiv1311.2595C}.

Brief phases of super-critical gas accretion that feed the MBH are also likely to feed intense bursts of star formation due to the high gas inflow rates. \cite{2013ApJ...772..112S} develops a simple feedback model that relates the star formation rate (SFR) to the MBH accretion rate. The model couples the two rates by incorporating outflows by jets or winds that induce pressure-enhanced star formation. 

We modify the model of \cite{2013ApJ...772..112S} for the case of super-critical MBH accretion, and discuss the implications in the following.  
  AGN triggering of star formation arises via pressure exerted on clouds, as jets and/or winds propagate into an inhomogeneous interstellar medium, leading to enhanced star formation rates \citep{wagner13}. In this case, the Kennicutt--Schmidt law for star formation rate becomes:

\begin{equation}
\dot \Sigma_\ast^{AGN}= {\epsilon_{SN}\over\sigma_d} \Sigma_{gas}\sqrt{\pi Gp_{AGN}\over f_g},
\end{equation}
with the AGN-induced pressure, $p_{AGN}\propto {L_E/(4\pi R^2 c)}$, $f_g$ is the gas fraction and $\epsilon_{SN}$ the efficiency of supernova feedback, which modulates the normalization of the Kennicutt--Schmidt law as an explanation of galactic star formation inefficiency. Averaging over the disk half-light radius, and expressing   the AGN-induced pressure as a function of the MBH growth rate, the AGN triggered star formation rate is related to the MBH accretion rate by
\begin{equation}
\dot M_\ast   ={\epsilon \zeta} f_{p} \dot M_{BH}, 
\end{equation}
where 
$\epsilon$ is the radiative efficiency, $\zeta=m_{SN}v c{/2 E_{SN}}$ is the SNe boost factor,  $f_{p}$ is the mechanical advantage factor of the wind or jet-driven bow shock \citep{wagner13}.
We define the supernova energy $E_{SN}=10^{51}E_{51} \rm ergs,$ the mass in stars formed per type II supernova 
$m_{SN}=150m_{150}\rm M_\odot, $ and the velocity at which SN-driven blast waves enter the cooling phase $v=400v_{400}  \rm km/s$. Plausible estimates for the parameters are $\zeta =180 m_{150}v_{400}{/ E_{51}}$, $f_{p}\sim 30$.   If indeed AGN outflows initially enhance star formation, such an effect must be  localized in the nucleus.  The star formation in obscured X-ray selected AGN is indeed significantly more concentrated than in galaxies 
with comparable high star formation rates \citep{2014ApJ...781L..34M}. 

For the radiatively-efficient Eddington-limited case $\epsilon\sim 0.2,$ and \cite{2013ApJ...772..112S} infers that, statistically\footnote{That statistically the ratio $\dot M_{BH}   / \dot M_\ast $ may be approximately constant over a wide range in $z$ was recognized in an early study by \cite{2008ApJ...679..118S}. An alternative way to tease out the statistical value of the ratio $\dot M_{BH}   / \dot M_\ast $ is to stack many galaxies, as in a recent analysis of SFR-selected  and X-ray stacked galaxies at $z\sim 0.5-2$ \citep{2012ApJ...753L..30M,2013ApJ...773....3C}.},  $\dot M_{BH}  \sim 10^{-3} \dot M_\ast $.  Over time, therefore, this leads to a ratio of MBH mass to stellar mass  $\sim 10^{-3}$, similar to that observed at low $z$ \citep[and references therein]{2013ARA&A..51..511K}.  If supercritical growth  is radiatively inefficient, then  $\epsilon  \ll 1,$ and therefore, for AGN-triggered star formation $\dot M_{\rm BH} \gg 10^{-3}  \dot M_\ast $.  MBHs that grew at least in part through radiatively inefficient supercritical accretion would therefore grow at a faster rate than the stellar mass of their host galaxies. Our model therefore naturally accommodates the suggestion that high redshift MBHs account for a higher fraction of the stellar mass of the host. For instance, \cite{2012MNRAS.420.3621T} suggest that the ratio of MBH mass to stellar mass is higher by an order of magnitude at $z\gsim 4$ than at $z=0$. 

\section{Observability}
Recent simulations \citep{2011ApJ...736....2O,2014MNRAS.439..503S,2013arXiv1312.6127M} suggest that MBHs accreting at  super-critical rates are most likely to be characterized by strongly collimated outflows or jets.   \cite{2011MNRAS.416..216V} and \cite{2013MNRAS.432.2818G} already suggested that the number density of jetted quasars approaches and possibly prevails over that of radio--quiet quasars at $z>4$,  based on the cosmic evolution indicated by the Swift all sky survey (Ajello et al. 2009). Their analysis takes into account that detection of sources with their jets pointing at us would be suppressed by a factor corresponding to the square of the jet opening angle:  the number of sources observed pointing at us is only a fraction $\sim\Omega^2$ of the the sources pointing in all directions. For example, for a funnel opening angle $\Omega\sim$ 0.1 rad \citep{2014MNRAS.439..503S} only one in 100 sources would be detected at high energies through their jets directed at us.  These jetted sources appear to be at best under--represented in the combined SDSS+FIRST survey (see Volonteri et al. 2011). Indeed, as suggested by \cite{2013arXiv1311.7147G}, detection of misaligned jets through radio lobes is hindered, as  at these high redshift radio emission may be suppressed by interaction with the cosmic microwave background.  The population of super-critical sources we predict may, therefore, explain the high-redshift population of jetted sources hinted to by the Swift all sky survey (Ajello et al. 2009).  \cite{2014arXiv1410.0364S} and \cite{2014MNRAS.440L.111G} indeed confirm the existence of an early peak ($z>4$) of activity in jetted AGN, in contrast to the main formation epoch of massive radioÐquiet quasars ($z\sim 2.5$). 

Another important point regarding  the detection of these sources is obscuration.  The gas density is very high and so accretion is obscured, regardless of any assumption on the accretion disk structure and properties.  A simple estimate on the column density can be obtained by assuming the gas free-fall rate at the Bondi radius\footnote{Very similar estimates are obtained by estimating the density of the gas accumulated in the inner pc by assuming an inflow rate, say 100 $\msun$/yr and a lifetime, $\sim10^4$ yr, or by taking an isothermal distribution as in section 3 and integrating from the capture radius onwards, as the integral is dominated by the mass distribution near $R_g$ \citep[cf.][]{1999MNRAS.308L..39F}.}. With $\dot M_{\rm inflow}\sim \sigma^3/{\rm G}$ and a free-fall timescale $t_{\rm ff}=(GM/R_B^3)^{-1/3}$ the column density can be written as:
\beq
N_H\sim f_{\rm ff} \frac{\sigma^4}{m_p G M}\sim f_{\rm ff} 10^{25} {\rm cm}^{-2} \sigma_{100}^4 M_7^{-1},
\label{NH}
\eeq
where $f_{\rm ff}$ is fraction of the free-fall rate characterizing the gas inflow. 

The density of the gas accumulated within the Bondi radius would be $\sim f_{\rm ff} 10^5 \msun {\rm pc}^{-3}\sigma_{100}^6 M_7^{-2}$, thereby masking most of the MBH growth to levels beyond Compton-thick  \citep[see][for a detailed model of observational signatures and spectral energy distribution of growing MBHs in high-z galaxies surrounded by dense gas envelopes]{2013MNRAS.433.1556Y}. Recent X-ray data indeed suggests that there may be a significant population of heavily obscured (Compton-thick) quasars \citep{2014ApJ...794..102S,2014ApJ...785...17L}.  \cite{2014arXiv1409.4413G} also suggest a novel technique to infer the presence of obscured AGN in high-z galaxies: high excitation CO transitions in the millimeter band, e.g. the CO(17--16) line, detectable by ALMA at  $z=7$ and beyond.  Such high excitation lines require the presence of high--energy photons ($>1$ keV, Schleicher et al. 2010) and would trace MBHs whose detection would otherwise be hindered by obscuration. Such millimeter diagnostics may help reveal the population of super-critical sources we predict. 

Given the inverse dependence on the MBH mass, there is a mass above which the opacity through the sphere $<$ 1 and the MBH accretion is ``unveiled''.  If we couple Eq.~\ref{NH} and Eq.~\ref{msigma} to highlight the dependence on MBH mass only, we obtain:
\beq
N_H \gtrsim 10^{24}\, f_{\rm ff}M_7^{-1/5} \left(\frac{\lambda}{0.01} \right)^{8/5}\  {\rm cm}^{-2} .
\label{NHMsigma}
\eeq

We can take two high-redshift quasars as an example to test the general validity of Eq.~\ref{NHMsigma}: ULASJ1120 and SDSSJ1148. For these quasars we can find in the literature estimates of their luminosity (hence, the accretion rate $\dot{M}_{\rm BH}$) and gas velocity dispersion $\sigma$ necessary to calculate the fraction of the free-fall rate, $f_{\rm ff}$ in Eq.~\ref{NHMsigma}, as well as the MBH mass \citep{Wang2010,2012ApJ...751L..25V}.  For ULASJ1120 the bolometric luminosity is $\sim 2\times 10^{47}$ erg/s, the MBH mass $\sim 2\times 10^9 \msun$, and the velocity dispersion $\sim 100$ km/s. For J1148 the bolometric luminosity is $\sim 7\times 10^{47}$ erg/s, the MBH mass $\sim 6\times 10^9 \msun$, and the velocity dispersion $\sim 160$ km/s.  By assuming a radiative efficiency of order 0.04-0.1, one obtains in both cases $f_{\rm ff}=\dot{M}_{\rm BH}G/\sigma^3 \sim 0.15-0.35$, and $N_H\sim $ few$ \times 10^{22}{\rm cm}^{-2}$, compatible with current limits  for ULASJ1120 \citep[$N_H< 10^{23}{\rm cm}^{-2}$,][]{2014A&A...563A..46M}. A smaller MBH in a galaxy with a similar inflow rate, therefore with a higher Eddington ratio, would instead be heavily obscured during its growth, but when the MBHs reach a sufficiently high mass, they would be bright at all wavelengths.

\section{Discussion and conclusions}
The possibility of widespread  super-critical accretion has often been advocated to describe either local \citep[e.g.,][]{Collin2002,WangNetzer2003,2004A&A...420L..23K} or high-redshift sources \citep[][]{VolonteriRees2005,2012MNRAS.425.2892W,2012MNRAS.424.1461L,2014ApJ...784L..38M,2014Sci...345.1330A}, or through recent revisions of Soltan's argument  \citep{2013arXiv1310.3833N}. 

We develop an analytical model to estimate the growth of super-critically accreting MBHs in high-redshift galaxies. We summarize our results below.
\begin{itemize}
\item We estimate that the duty cycle of MBH growth can be as low as $\sim 0.01$, rather than the $\sim$ unity value required if accretion proceeds sub-critically.
\item The trapping of radiation in the disk, coupled with the angular momentum of the gas that ends in the accretion disk, provide a natural ``clock" for the properties of the accretion episode and the final mass of the MBH.  The lower the gas angular momentum, the longer accretion can continue, the higher the final MBH mass.
\item Linking the inflow rate to the Bondi rate, we are able to obtain a relation between the MBH mass and the velocity dispersion of the gas in the host galaxy at the end of the super-critical episode. In galaxies with low-angular momentum gas the MBH can get to a higher mass at fixed gas velocity dispersion, with respect to the normalization one would extrapolate from local galaxies.
\item We relate the MBH growth to the star formation rate in the galaxy. The model couples the two rates through outflows by jets or winds that induce pressure-enhanced star formation. Our model predicts that the ratio of MBH accretion rate to star formation rate is  $10^{-2}$ or higher, naturally leading to a ratio of MBH to stellar mass higher than the ``canonical" value of $\sim 10^{-3}$, in agreement with current observations at early epochs. 
\end{itemize}

Finally, we note that in the case of super-critical accretion, the radiative efficiency is not determined by MBH spin, thus the expected spin-up caused by prolonged accretion phases  would not hinder MBH growth through a high radiative efficiency.

\acknowledgements 
We thank M. Begelman, F. Governato and A. Babul for insightful comments.   We are grateful to Y. Dubois, J. Devriendt and C. Pichon for allowing us to use results from the  Horizon-AGN simulation. 
MV acknowledges funding support for this research from NASA, through Award Number ATP NNX10AC84G; from SAO, through Award Number TM1-12007X, from NSF, through Award Number AST 1107675, and from a Marie Curie FP7-Reintegration-Grants within the 7th European Community Framework Programme (PCIG10-GA-2011-303609).  The research of JS has been supported at IAP by  the ERC project  267117 (DARK)  hosted by Universit\'e Pierre et Marie Curie - Paris 6  and at JHU by NSF grant OIA-1124403.


\begin{thebibliography}{99}

\expandafter\ifx\csname natexlab\endcsname\relax\def\natexlab#1{#1}\fi

\bibitem[Abramowicz et al.(1988)]{1988ApJ...332..646A} Abramowicz, M.~A., 
Czerny, B., Lasota, J.~P., \& Szuszkiewicz, E.\ 1988, \apj, 332, 646 

\bibitem[Abramowicz 
\& Lasota(1995)]{1995ComAp..18..141A} Abramowicz, M.~A., \& Lasota, J.-P.\ 1995, Comments on Astrophysics, 18, 141 

\bibitem[{{Ajello}(2009)}]{Ajello2009} Ajello M., Costamante L., Sambruna R.M., et al., 2009, ApJ, 699, 603

\bibitem[Alexander 
\& Natarajan(2014)]{2014Sci...345.1330A} Alexander, T., \& Natarajan, P.\ 2014, Science, 345, 1330 

\bibitem[Antonuccio-Delogu 
\& Silk(2010)]{2010MNRAS.405.1303A} Antonuccio-Delogu, V., \& Silk, J.\ 2010, \mnras, 405, 1303 


\bibitem[Begelman(1979)]{1979MNRAS.187..237B} Begelman, M.~C.\ 1979, 
\mnras, 187, 237 


\bibitem[Begelman(2012a)]{2012MNRAS.420.2912B} Begelman, M.~C.\ 2012a, 
\mnras, 420, 2912 

\bibitem[Begelman(2012b)]{2012ApJ...749L...3B} Begelman, M.~C.\ 2012b, \apjl, 
749, L3 

\bibitem[Begelman et al.(2006)]{2006MNRAS.370..289B} Begelman, M.~C., 
Volonteri, M., \& Rees, M.~J.\ 2006, \mnras, 370, 289 

\bibitem[Blackburne et al.(2011)]{2011ApJ...729...34B} Blackburne, J.~A., 
Pooley, D., Rappaport, S., \& Schechter, P.~L.\ 2011, \apj, 729, 34 

\bibitem[Blandford 
\& Begelman(1999)]{1999MNRAS.303L...1B} Blandford, R.~D., \& Begelman, M.~C.\ 1999, \mnras, 303, L1 

\bibitem[Blandford 
\& Begelman(2004)]{2004MNRAS.349...68B} Blandford, R.~D., \& Begelman, M.~C.\ 2004, \mnras, 349, 68 

\bibitem[Bondi 
\& Hoyle(1944)]{1944MNRAS.104..273B} Bondi, H., \& Hoyle, F.\ 1944, \mnras, 104, 273 

\bibitem[{{Bongiorno} {et~al.}(2012)
{Bongiorno}, {Merloni}, {Brusa},
  {Magnelli}, {Salvato}, {Mignoli}, {Zamorani}, {Fiore}, {Rosario}, {Mainieri},
  {Hao}, {Comastri}, {Vignali}, {Balestra}, {Bardelli}, {Berta}, {Civano},
  {Kampczyk}, {Le Floc'h}, {Lusso}, {Lutz}, {Pozzetti}, {Pozzi}, {Riguccini},
  {Shankar}, \& {Silverman}}]{Bongiorno2012}
  Bongiorno, A. et al. 
  2012, \mnras, 427, 3103


\bibitem[Chen et al.(2013)]{2013ApJ...773....3C} Chen, C.-T.~J., Hickox, 
R.~C., Alberts, S., et al.\ 2013, \apj, 773, 3 


\bibitem[Cicone et al.(2013)]{2013arXiv1311.2595C} Cicone, C., Maiolino, R., Sturm, E., et al.\ 2014, \aap, 562, A21

\bibitem[Cielo et al.(2013)]{2013arXiv1311.5562C} Cielo, S., 
Antonuccio-Delogu, V., Macci{\`o}, A.~V., Romeo, A.~D., 
\& Silk, J.\ 2014, \mnras, 439, 2903 

\bibitem[{{Collin} {et~al.}(2002){Collin}, {Boisson}, {Mouchet}, {Dumont},
  {Coup{\'e}}, {Porquet}, \& {Rokaki}}]{Collin2002}
{Collin}, S. , {Boisson}, C., {Mouchet}, M., {Dumont}, A.-M., {Coup{\'e}}, S., {Porquet}, D., \& {Rokaki}, E. 
  2002, \aap, 388, 771


\bibitem[Costa et al.(2014)]{2014MNRAS.439.2146C} Costa, T., Sijacki, D., 
Trenti, M., \& Haehnelt, M.~G.\ 2014, \mnras, 439, 2146 


\bibitem[Coughlin 
\& Begelman(2013)]{2013arXiv1312.5314C} Coughlin, E.~R., \& Begelman, M.~C.\ 2014, \apj, 781, 82 


\bibitem[{{De Rosa} {et~al.}(2013){De Rosa}, {Venemans}, {Decarli}, {Gennaro},
  {Simcoe}, {Dietrich}, {Peterson}, {Walter}, {Frank}, {McMahon}, {Hewett},
  {Mortlock}, \& {Simpson}}]{DeRosa2013}
De Rosa, G., Venemans, 
B.~P., Decarli, R., et al.\ 2014, \apj, 790, 145 


\bibitem[{{Dubois} {et~al.}(2013){Dubois}, {Pichon}, {Devriendt}, {Silk},
  {Haehnelt}, {Kimm}, \& {Slyz}}]{Dubois2013}
{Dubois}, Y. , {Pichon}, C., {Devriendt}, J., {Silk}, J., {Haehnelt}, M., {Kimm}, T., \& {Slyz}, A. 
  2013, \mnras, 428, 2885

\bibitem[Dubois et al.(2014)]{2014MNRAS.444.1453D} Dubois, Y., Pichon, C., 
Welker, C., et al.\ 2014, \mnras, 444, 1453 

\bibitem[Edelson et al.(2015)]{2015arXiv150105951E} Edelson, R., Gelbord, 
J.~M., Horne, K., et al.\ 2015, arXiv:1501.05951 

\bibitem[Fabian(1999)]{1999MNRAS.308L..39F} Fabian, A.~C.\ 1999, \mnras, 
308, L39 

\bibitem[Feng et al.(2014)]{2014MNRAS.440.1865F} Feng, Y., Di Matteo, T., 
Croft, R., \& Khandai, N.\ 2014, \mnras, 440, 1865 

\bibitem[Gabor 
\& Bournaud(2013)]{2013MNRAS.434..606G} Gabor, J.~M., \& Bournaud, F.\ 2013, \mnras, 434, 606 

\bibitem[Gaibler et al.(2012)]{2012MNRAS.425..438G} Gaibler, V., Khochfar, 
S., Krause, M., \& Silk, J.\ 2012, \mnras, 425, 438 

\bibitem[Gallerani et al.(2014)]{2014arXiv1409.4413G} Gallerani, S., 
Ferrara, A., Neri, R., \& Maiolino, R.\ 2014, arXiv:1409.4413 


\bibitem[Ghisellini et al.(2013)]{2013MNRAS.432.2818G} Ghisellini, G., 
Haardt, F., Della Ceca, R., Volonteri, M., 
\& Sbarrato, T.\ 2013, \mnras, 432, 2818 

\bibitem[Ghisellini et al.(2013)]{2013arXiv1311.7147G} Ghisellini, G., 
Celotti, A., Tavecchio, F., Haardt, F., 
\& Sbarrato, T.\ 2014, \mnras, 438, 2694 

\bibitem[Ghisellini et al.(2014)]{2014MNRAS.440L.111G} Ghisellini, G., 
Sbarrato, T., Tagliaferri, G., et al.\ 2014, \mnras, 440, L111 


\bibitem[\protect\citeauthoryear{{Goodman} \& {Tan}}{{Goodman} \&
  {Tan}}{2004}]{goodman&tan04}
{Goodman} J.,  {Tan} J.~C.,  2004, \apj, 608, 108


\bibitem[Harrison et al.(2014)]{2014arXiv1403.3086H} Harrison, C.~M., 
Alexander, D.~M., Mullaney, J.~R., 
\& Swinbank, A.~M.\ 2014, \mnras, 441, 3306 

\bibitem[Hirschmann et al.(2014)]{2014MNRAS.442.2304H} Hirschmann, M., 
Dolag, K., Saro, A., et al.\ 2014, \mnras, 442, 2304 


\bibitem[Inayoshi 
\& Haiman(2014)]{2014arXiv1406.5058I} Inayoshi, K., \& Haiman, Z.\ 2014, arXiv:1406.5058 

\bibitem[Jiang et al.(2014)]{2014arXiv1410.0678J} Jiang, Y.-F., Stone, 
J.~M., \& Davis, S.~W.\ 2014, arXiv:1410.0678 

\bibitem[{{Johnson} {et~al.}(2012){Johnson}, {Whalen}, {Fryer}, \&
  {Li}}]{Johnson2012}
{Johnson}, J.~L. , {Whalen}, D.~J., {Fryer}, C.~L., \& {Li}, H. 
2012, \apj, 750,
  66
  
  \bibitem[Kawaguchi et 
al.(2004)]{2004A&A...415...47K} Kawaguchi, T., Pierens, A., \& Hur{\'e}, J.-M.\ 2004, \aap, 415, 47 

  \bibitem[Kawaguchi et 
al.(2004)]{2004A&A...420L..23K} Kawaguchi, T., Aoki, K., Ohta, K., \& Collin, S.\ 2004, \aap, 420, L23 

\bibitem[Kolokythas et al.(2014)]{2014arXiv1402.5109K} Kolokythas, K., 
O'Sullivan, E., Raychaudhury, S., Ishwara-Chandra, C.~H., 
\& Kantharia, N.\ 2014, arXiv:1402.5109 

\bibitem[Kormendy 
\& Ho(2013)]{2013ARA&A..51..511K} Kormendy, J., \& Ho, L.~C.\ 2013, \araa, 51, 511 

\bibitem[Krumholz 
\& Tan(2007)]{2007ApJ...654..304K} Krumholz, M.~R., \& Tan, J.~C.\ 2007, \apj, 654, 304 

\bibitem[Laing 
\& Bridle(2012)]{2012MNRAS.424.1149L} Laing, R.~A., \& Bridle, A.~H.\ 2012, \mnras, 424, 1149 

\bibitem[Lansbury et al.(2014)]{2014ApJ...785...17L} Lansbury, G.~B., 
Alexander, D.~M., Del Moro, A., et al.\ 2014, \apj, 785, 17 

 \bibitem[Laor 
\& Netzer(1989)]{1989MNRAS.238..897L} Laor, A., \& Netzer, H.\ 1989, \mnras, 238, 897  
  
\bibitem[Li et al.(2007)]{2007ApJ...665..187L} Li, Y., Hernquist, L., 
Robertson, B., et al.\ 2007, \apj, 665, 187 


\bibitem[{{Li}(2012)}]{2012MNRAS.424.1461L}
{Li}, L.-X. 2012, \mnras, 424, 1461

\bibitem[Madau et al.(2014)]{2014ApJ...784L..38M} Madau, P., Haardt, F., 
\& Dotti, M.\ 2014, \apjl, 784, L38 


\bibitem[Matsuoka et al.(2013)]{2013arXiv1312.2417M} Matsuoka, Y., Strauss, 
M.~A., Price, T.~N., III, \& DiDonato, M.~S.\ 2014, \apj, 780, 162 

\bibitem[McConnell 
\& Ma(2013)]{2013ApJ...764..184M} McConnell, N.~J., \& Ma, C.-P.\ 2013, \apj, 764, 184 

\bibitem[{{McKinney} {et~al.}(2013){McKinney}, {Tchekhovskoy}, {Sadowski}, \&
  {Narayan}}]{2013arXiv1312.6127M}
McKinney, J.~C.,  Tchekhovskoy, A., Sadowski, A., \& Narayan, R.\ 2014, \mnras, 441, 3177 

\bibitem[{{Mineshige} {et~al.}(2000){Mineshige}, {Kawaguchi}, {Takeuchi}, \&
  {Hayashida}}]{Mineshige2000}
{Mineshige}, S.  {Kawaguchi}, T., {Takeuchi}, M., \& {Hayashida}, K. 
2000,
  \pasj, 52, 499

\bibitem[Moretti et 
al.(2014)]{2014A&A...563A..46M} Moretti, A., Ballo, L., Braito, V., et al.\ 2014, \aap, 563, A46 

\bibitem[Morganti et 
al.(2013)]{2013A&A...552L...4M} Morganti, R., Frieswijk, W., Oonk, R.~J.~B., Oosterloo, T., \& Tadhunter, C.\ 2013, \aap, 552, L4 


\bibitem[{{Mullaney} {et~al.}(2012){Mullaney}, {Daddi}, {B{\'e}thermin},
  {Elbaz}, {Juneau}, {Pannella}, {Sargent}, {Alexander}, \&
  {Hickox}}]{2012ApJ...753L..30M}
{Mullaney}, J.~R. , {Daddi}, E., {B{\'e}thermin}, M., {Elbaz}, D., {Juneau}, S.,  {Pannella}, M., {Sargent}, M.~T., {Alexander}, D.~M., \& {Hickox}, R.~C.
  2012, \apjl, 753, L30

\bibitem[Mushotzky et al.(2014)]{2014ApJ...781L..34M} Mushotzky, R.~F., 
Shimizu, T.~T., Mel{\'e}ndez, M., \& Koss, M.\ 2014, \apjl, 781, L34 

\bibitem[Narayan 
\& Yi(1995)]{1995ApJ...452..710N} Narayan, R., \& Yi, I.\ 1995, \apj, 452, 710 

\bibitem[Netzer 
\& Trakhtenbrot(2013)]{2013MNRAS.tmp.2900N} Netzer, H., \& Trakhtenbrot, B.\ 2013, \mnras, 2900 


\bibitem[{{Netzer} {et~al.}(2013){Netzer}, {Mor}, {Trakhtenbrot}, {Shemmer}, \&
  {Lira}}]{2013arXiv1308.0012N}
Netzer, H., Mor, R., Trakhtenbrot, B., Shemmer, O., \& Lira, P.\ 2014, \apj, 791, 34 

\bibitem[{{Novak}(2013)}]{2013arXiv1310.3833N}
{Novak}, G.~S. 2013, ArXiv e-prints, arXiv:1310.3833

\bibitem[Ohsuga(2007)]{2007ApJ...659..205O} Ohsuga, K.\ 2007, \apj, 659, 
205

\bibitem[Ohsuga et al.(2002)]{2002ApJ...574..315O} Ohsuga, K., Mineshige, 
S., Mori, M., \& Umemura, M.\ 2002, \apj, 574, 315 

\bibitem[Ohsuga et al.(2005)]{2005ApJ...628..368O} Ohsuga, K., Mori, M., 
Nakamoto, T., \& Mineshige, S.\ 2005, \apj, 628, 368 


\bibitem[Ohsuga \& Mineshige(2011)]{2011ApJ...736....2O} Ohsuga, K., \& Mineshige, S.\ 2011, \apj, 736, 2 

\bibitem[Park 
\& Ricotti(2012)]{2012ApJ...747....9P} Park, K., \& Ricotti, M.\ 2012, \apj, 747, 9 

\bibitem[{{Paczynski}(1982)}]{Paczynski1982}
{Paczynski}, B. 1982, Mitteilungen der Astronomischen Gesellschaft Hamburg, 57,
  27
  
\bibitem[Poutanen et al.(2007)]{2007MNRAS.377.1187P} Poutanen, J., 
Lipunova, G., Fabrika, S., Butkevich, A.~G., 
\& Abolmasov, P.\ 2007, \mnras, 377, 1187 

\bibitem[{{Rafferty} {et~al.}(2011){Rafferty}, {Brandt}, {Alexander}, {Xue},
  {Bauer}, {Lehmer}, {Luo}, \& {Papovich}}]{2011ApJ...742....3R}
{Rafferty}, D.~A. , {Brandt}, W.~N., {Alexander}, D.~M., {Xue}, Y.~Q., {Bauer}, F.~E., {Lehmer}, B.~D., {Luo}, B., \& {Papovich}, C. 
  2011, \apj, 742, 3
 
   
\bibitem[Rashed et 
al.(2013)]{2013A&A...558A...5R} Rashed, Y.~E., Zuther, J., Eckart, A., et al.\ 2013, \aap, 558, A5 

 \bibitem[Rocca-Volmerange et al.(2013)]{2013MNRAS.429.2780R} 
Rocca-Volmerange, B., Drouart, G., De Breuck, C., et al.\ 2013, \mnras, 
429, 2780 

 
 \bibitem[Rosario et 
al.(2012)]{2012A&A...545A..45R} Rosario, D.~J., Santini, P., Lutz, D., et al.\ 2012, \aap, 545, A45 

\bibitem[Rosario et 
al.(2013)]{2013A&A...560A..72R} Rosario, D.~J., Trakhtenbrot, B., Lutz, D., et al.\ 2013, \aap, 560, A72 


\bibitem[S{\c a}dowski et al.(2014)]{2014MNRAS.439..503S} S{\c a}dowski, 
A., Narayan, R., McKinney, J.~C., 
\& Tchekhovskoy, A.\ 2014, \mnras, 439, 503 

\bibitem[Sbarrato et al.(2014)]{2014arXiv1410.0364S} Sbarrato, T., 
Ghisellini, G., Tagliaferri, G., et al.\ 2014, arXiv:1410.0364 

\bibitem[Schleicher et 
al.(2010)]{2010A&A...513A...7S} Schleicher, D.~R.~G., Spaans, M., \& Klessen, R.~S.\ 2010, \aap, 513, A7 


\bibitem[\protect\citeauthoryear{{Shakura} \& {Sunyaev}}{{Shakura} \&
  {Sunyaev}}{1973}]{shakura&sunyaev73}
{Shakura} N.~I.,  {Sunyaev} R.~A.,  1973, \aap, 24, 337

\bibitem[Shlosman et al.(1989)]{1989Natur.338...45S} Shlosman, I., Frank, 
J., \& Begelman, M.~C.\ 1989, \nat, 338, 45 

\bibitem[Silk(2013)]{2013ApJ...772..112S} Silk, J.\ 2013, \apj, 772, 112 

\bibitem[Silverman et al.(2008)]{2008ApJ...679..118S} Silverman, J.~D., 
Green, P.~J., Barkhouse, W.~A., et al.\ 2008, \apj, 679, 118 


\bibitem[Spaans 
\& Silk(2006)]{2006ApJ...652..902S} Spaans, M., \& Silk, J.\ 2006, \apj, 652, 902 


\bibitem[Stern et al.(2014)]{2014ApJ...794..102S} Stern, D., Lansbury, 
G.~B., Assef, R.~J., et al.\ 2014, \apj, 794, 102


\bibitem[Sturm et al.(2011)]{2011ApJ...733L..16S} Sturm, E., 
Gonz{\'a}lez-Alfonso, E., Veilleux, S., et al.\ 2011, \apjl, 733, L16 


\bibitem[{{Symeonidis} {et~al.}(2011){Symeonidis}, {Georgakakis}, {Seymour},
  {Auld}, {Bock}, {Brisbin}, {Buat}, {Burgarella}, {Chanial}, {Clements},
  {Cooray}, {Eales}, {Farrah}, {Franceschini}, {Glenn}, {Griffin},
  {Hatziminaoglou}, {Ibar}, {Ivison}, {Mortier}, {Oliver}, {Page},
  {Papageorgiou}, {Pearson}, {P{\'e}rez-Fournon}, {Pohlen}, {Rawlings},
  {Raymond}, {Rodighiero}, {Roseboom}, {Rowan-Robinson}, {Scott}, {Smith},
  {Tugwell}, {Vaccari}, {Vieira}, {Vigroux}, {Wang}, \&
  {Wright}}]{2011MNRAS.417.2239S}
{Symeonidis}, M. et al.
 2011, \mnras, 417, 2239

\bibitem[Tanaka(2014)]{2014CQGra..31x4005T} Tanaka, T.~L.\ 2014, Classical 
and Quantum Gravity, 31, 244005 

\bibitem[Takeuchi et al.(2009)]{2009PASJ...61..783T} Takeuchi, S., 
Mineshige, S., \& Ohsuga, K.\ 2009, \pasj, 61, 783 


\bibitem[Targett et al.(2012)]{2012MNRAS.420.3621T} Targett, T.~A., Dunlop, 
J.~S., \& McLure, R.~J.\ 2012, \mnras, 420, 3621 

\bibitem[Venemans et al.(2012)]{2012ApJ...751L..25V} Venemans, B.~P., 
McMahon, R.~G., Walter, F., et al.\ 2012, \apjl, 751, L25 

\bibitem[Vernaleo 
\& Reynolds(2006)]{2006ApJ...645...83V} Vernaleo, J.~C., \& Reynolds, C.~S.\ 2006, \apj, 645, 83 

\bibitem[{{Volonteri} \& {Rees}(2005)}]{VolonteriRees2005}
{Volonteri}, M. \& {Rees}, M.~J. 2005, {ApJ}, 633, 624

\bibitem[Volonteri et al.(2011)]{2011MNRAS.416..216V} Volonteri, M., Haardt, F., Ghisellini, G., \& Della Ceca, R.\ 
2011, \mnras, 416, 216 


\bibitem[Wagner et al.(2012)]{2012ApJ...757..136W} Wagner, A.~Y., Bicknell, 
G.~V., \& Umemura, M.\ 2012, \apj, 757, 136 


\bibitem[Wagner et al.(2013)]{wagner13} Wagner, A., Umemura, 
M., \& Bicknell, G.\ 2013, \apjl, 763, L18 


\bibitem[{{Wang} {et~al.}(2006){Wang}, {Chen}, \& {Hu}}]{Wang2006a}
{Wang}, J.-M., {Chen}, Y.-M., \& {Hu}, C. 2006, \apjl, 637, L85

\bibitem[{{Wang} \& {Netzer}(2003)}]{WangNetzer2003}
{Wang}, J.-M. \& {Netzer}, H. 2003, \aap, 398, 927

\bibitem[{{Wang} {et~al.}(2010){Wang}, {Carilli}, {Neri}, {Riechers}, {Wagg},
  {Walter}, {Bertoldi}, {Menten}, {Omont}, {Cox}, \& {Fan}}]{Wang2010}
{Wang}, R. , {Carilli}, C.~L., {Neri}, R., {Riechers}, D.~A., {Wagg}, J.,
  {Walter}, F., {Bertoldi}, F., {Menten}, K.~M., {Omont}, A., {Cox}, P., \&
  {Fan}, X. 
  2010, ApJ, 714, 699

\bibitem[{{Willott} {et~al.}(2013){Willott}, {Omont}, \&
  {Bergeron}}]{2013ApJ...770...13W}
{Willott}, C.~J., {Omont}, A., \& {Bergeron}, J. 2013, \apj, 770, 13

\bibitem[{{Wyithe} \& {Loeb}(2012)}]{2012MNRAS.425.2892W}
{Wyithe}, J.~S.~B. \& {Loeb}, A. 2012, \mnras, 425, 2892

\bibitem[Yue et al.(2013)]{2013MNRAS.433.1556Y} Yue, B., Ferrara, A., 
Salvaterra, R., Xu, Y., \& Chen, X.\ 2013, \mnras, 433, 1556 


\bibitem[Zinn et al.(2013)]{2013ApJ...774...66Z} Zinn, P.-C., Middelberg, 
E., Norris, R.~P., \& Dettmar, R.-J.\ 2013, \apj, 774, 66 


\end{thebibliography}

\end{document}